\documentclass[aps,prl,twocolumn,superscriptaddress,showpacs,preprintnumbers,amsmath,amssymb]{revtex4}
\usepackage{graphicx} 
\usepackage{dcolumn}  

\begin{document}


\preprint{\vbox{ \hbox{   }
                 \hbox{  }
                 \hbox{BELLE-CONF-0576}
}}

\title{ \quad\\[0.5cm]  \ \\ \ \\
Observation of a $\chi'_{c2}$ candidiate in\\ 
$\gamma \gamma \to D\bar{D}$ production in Belle }

\affiliation{Aomori University, Aomori}
\affiliation{Budker Institute of Nuclear Physics, Novosibirsk}
\affiliation{Chiba University, Chiba}
\affiliation{Chonnam National University, Kwangju}
\affiliation{University of Cincinnati, Cincinnati, Ohio 45221}
\affiliation{University of Frankfurt, Frankfurt}
\affiliation{Gyeongsang National University, Chinju}
\affiliation{University of Hawaii, Honolulu, Hawaii 96822}
\affiliation{High Energy Accelerator Research Organization (KEK), Tsukuba}
\affiliation{Hiroshima Institute of Technology, Hiroshima}
\affiliation{Institute of High Energy Physics, Chinese Academy of Sciences, Beijing}
\affiliation{Institute of High Energy Physics, Vienna}
\affiliation{Institute for Theoretical and Experimental Physics, Moscow}
\affiliation{J. Stefan Institute, Ljubljana}
\affiliation{Kanagawa University, Yokohama}
\affiliation{Korea University, Seoul}
\affiliation{Kyoto University, Kyoto}
\affiliation{Kyungpook National University, Taegu}
\affiliation{Swiss Federal Institute of Technology of Lausanne, EPFL, Lausanne}
\affiliation{University of Ljubljana, Ljubljana}
\affiliation{University of Maribor, Maribor}
\affiliation{University of Melbourne, Victoria}
\affiliation{Nagoya University, Nagoya}
\affiliation{Nara Women's University, Nara}
\affiliation{National Central University, Chung-li}
\affiliation{National Kaohsiung Normal University, Kaohsiung}
\affiliation{National United University, Miao Li}
\affiliation{Department of Physics, National Taiwan University, Taipei}
\affiliation{H. Niewodniczanski Institute of Nuclear Physics, Krakow}
\affiliation{Nippon Dental University, Niigata}
\affiliation{Niigata University, Niigata}
\affiliation{Nova Gorica Polytechnic, Nova Gorica}
\affiliation{Osaka City University, Osaka}
\affiliation{Osaka University, Osaka}
\affiliation{Panjab University, Chandigarh}
\affiliation{Peking University, Beijing}
\affiliation{Princeton University, Princeton, New Jersey 08544}
\affiliation{RIKEN BNL Research Center, Upton, New York 11973}
\affiliation{Saga University, Saga}
\affiliation{University of Science and Technology of China, Hefei}
\affiliation{Seoul National University, Seoul}
\affiliation{Shinshu University, Nagano}
\affiliation{Sungkyunkwan University, Suwon}
\affiliation{University of Sydney, Sydney NSW}
\affiliation{Tata Institute of Fundamental Research, Bombay}
\affiliation{Toho University, Funabashi}
\affiliation{Tohoku Gakuin University, Tagajo}
\affiliation{Tohoku University, Sendai}
\affiliation{Department of Physics, University of Tokyo, Tokyo}
\affiliation{Tokyo Institute of Technology, Tokyo}
\affiliation{Tokyo Metropolitan University, Tokyo}
\affiliation{Tokyo University of Agriculture and Technology, Tokyo}
\affiliation{Toyama National College of Maritime Technology, Toyama}
\affiliation{University of Tsukuba, Tsukuba}
\affiliation{Utkal University, Bhubaneswer}
\affiliation{Virginia Polytechnic Institute and State University, Blacksburg, Virginia 24061}
\affiliation{Yonsei University, Seoul}
  \author{K.~Abe}\affiliation{High Energy Accelerator Research Organization (KEK), Tsukuba} 
  \author{K.~Abe}\affiliation{Tohoku Gakuin University, Tagajo} 
  \author{I.~Adachi}\affiliation{High Energy Accelerator Research Organization (KEK), Tsukuba} 
  \author{H.~Aihara}\affiliation{Department of Physics, University of Tokyo, Tokyo} 
  \author{K.~Aoki}\affiliation{Nagoya University, Nagoya} 
  \author{K.~Arinstein}\affiliation{Budker Institute of Nuclear Physics, Novosibirsk} 
  \author{Y.~Asano}\affiliation{University of Tsukuba, Tsukuba} 
  \author{T.~Aso}\affiliation{Toyama National College of Maritime Technology, Toyama} 
  \author{V.~Aulchenko}\affiliation{Budker Institute of Nuclear Physics, Novosibirsk} 
  \author{T.~Aushev}\affiliation{Institute for Theoretical and Experimental Physics, Moscow} 
% ...
\collaboration{The Belle Collaboration}
\noaffiliation

\begin{abstract}
We report on a search for the 
production of new resonance states
in the process $\gamma \gamma \to D \bar{D}$.
A candidate
$C$-even charmonium state is observed
in the vicinity of 3.93~GeV/$c^2$.  The production
rate and the angular distribution in the $\gamma\gamma$
center-of-mass frame suggest that this state is the
previously unobserved $\chi'_{c2}$, the $2^3P_2$
charmonium state.
\end{abstract}


\maketitle

\tighten

{\renewcommand{\thefootnote}{\fnsymbol{footnote}}}
\setcounter{footnote}{0}

\section{Introduction}
  The masses and other properties of 
the radial-ground and radially excited states
of charmonium provide valuable input to QCD models
that describe heavy quarkonium systems.  To date, 
radial excitation states of charmonium 
have been found only for
the $^{2S+1}L_J =$ $^3S_1$ states ($\psi$) and
$^1S_0$ states ($\eta_c$).  No radially excited
$^3P_J$ states ($\chi_{cJ}$) have yet been
found, even though the three radial ground states 
have been already well established.

The first radially excited $\chi_{cJ}$ states
are predicted to have masses between
3.9 and 4.0~GeV/$c^2$~\cite{godfrey,eichten}, which is
considerably above $D\bar{D}$ 
threshold.  If the masses of these states lie between
the $D\bar{D}$ and $D^*\bar{D^*}$ thresholds, the 
$\chi_{c0}(2P)\ (\chi'_{c0})$  and $\chi_{c2}(2P)\ (\chi'_{c2})$
are expected to decay primarily into $D\bar{D}$, although 
the $\chi'_{c2}$ could also decay to $D\bar{D}^*$ if it is
energetically allowed.
(The inclusion of charge-conjugated reactions
is implied throughout this paper.)
Recently,  two new charmonium-like states in this mass region, 
the $X(3940)$~\cite{belledccbar} and $Y(3940)$~\cite{y3940}, 
were reported by Belle.  Decays of either of these states
to $D\bar{D}$ have not been observed~\cite{belledccbar}.

In this paper we report on a search for the 
$\chi'_{cJ}$ ($J=0$ or 2) states
and other $C$-even charmonium states
in the mass range of 3.73 - 4.3~GeV/$c^2$
produced via the $\gamma \gamma \to D\bar{D}$ process.
The results presented here are preliminary.

\section{Data and detector}
The analysis uses data recorded in
the Belle detector at the KEKB asymmetric $e^+e^-$ 
collider~\cite{kekb}.
The data sample corresponds to 
an integrated luminosity of 280~fb$^{-1}$,
accumulated on the 
$\Upsilon(4S)$ resonance $(\sqrt{s} = 10.58~{\rm GeV})$
and 60~MeV below the resonance.
Since the beam energy dependence of 
two-photon processes is small, we combine
both samples.  We study the two-photon process
$e^+e^-\rightarrow e^+e^- D\bar{D}$
in the ``no-tag'' mode, {\em i.e.} where  
neither the final-state electron nor positron is detected. 
We restrict the virtuality of the incident photons to 
be small by imposing a strict requirement on the
transverse-momentum balance of the final-state 
hadronic system with respect to the beam axis  .

A comprehensive description of the Belle detector is
given elsewhere~\cite{belle}. We mention here only the
detector components essential for the present measurement.

Charged tracks are reconstructed from hit information in a central
drift chamber (CDC) located in a uniform 1.5~T solenoidal magnetic field.
The $z$ axis of the detector and the solenoid are along the positron beam,
with the positrons moving in the $-z$ direction.  The CDC measures the
longitudinal and transverse momentum components (along the $z$ axis and
in the $r\varphi$ plane, respectively).
Track trajectory coordinates near the
collision point are measured by a
silicon vertex detector (SVD).  Photon detection and
energy measurements are provided by a CsI(Tl) electromagnetic
calorimeter (ECL).   Species of charged hadron are
identified by means of information from
time-of-flight counters (TOF) and a silica-aerogel Cherenkov
counters (ACC).  The ACC provides separation between kaons 
and pions for momenta above
1.2~GeV/$c$.  The TOF system consists of a barrel of 128 
plastic scintillation counters, and is effective for $K/\pi$ 
separation for tracks with momenta below 1.2~GeV/$c$. 
Low energy kaons are also identified by specific
ionization ($dE/dx$) measurements in the CDC.

Kaon candidates are separated from pions based on normalized kaon and pion 
likelihood functions obtained from the particle identification system 
($L_K$ and $L_{\pi}$, respectively) 
with a criterion, $L_K/(L_K+L_{\pi})>0.8$.
All tracks that are not identified as kaons are treated as
pions. 

Signal candidates are triggered by a variety
of track-triggers that require two or more CDC 
tracks with associated TOF hits, ECL clusters or a minimum
sum of energy in the ECL.  For the four and six charged
track topologies used in this analysis,
the trigger conditions are complementary to each other
and, in combination, provide a
high trigger efficiency ($\sim$ 95\%).

\section{Event selection}
We search for exclusive $D\bar{D}$ production in
the following four combinations of decays:
\begin{eqnarray}
\gamma\gamma &\to& D^0\bar{D}^0,\ D^0 \to K^-\pi^+,\ \bar{D}^0 \to K^+\pi^-\ \
\ \ \ \ \ \ \ \ {\rm (N4)}, \nonumber \\
\gamma\gamma &\to& D^0\bar{D}^0,\ D^0 \to K^-\pi^+,\ \bar{D}^0 \to K^+\pi^-\pi^0
\ \ \ \ \ \ \ {\rm (N5)}, \nonumber \\
\gamma\gamma &\to& D^0\bar{D}^0,\ D^0 \to K^-\pi^+,\ \bar{D}^0 \to K^+\pi^-\pi^+
\pi^-\ \ \ {\rm (N6)}, \nonumber \\
\gamma\gamma &\to& D^+D^-,\ D^+ \to K^-\pi^+\pi^+,\ D^- \to K^+\pi^-\pi^-\ {\rm (C6)}. \nonumber
\end{eqnarray}
The symbols in parentheses are used to designate
each of the final states.
For the four-prong processes (N4 and N5) the selection criteria are:
four charged tracks, each one with (L) a transverse momentum 
in the laboratory frame 
of $p_t >0.1$~GeV/$c$ and a distance of closest approach 
to the nominal collision point of $|dr|<5$~cm and $|dz|<5~$cm;
the absolute value of the average $dz$ for all of the tracks,
$|\overline{dz}|<3$~cm;
two or more of the four tracks must have (S) $p_t >0.4$~GeV/$c$,
$|dr|<1$~cm, and $-0.8660 < \cos \theta <+0.9563$, where $\theta$ is
the laboratory frame polar angle; 
no photon clusters with an energy greater than 400~MeV; 
the charged track system consists of a 
$K^+K^-\pi^+\pi^-$ combination;
the larger of the two neutral $K\pi$ invariant masses 
should lie within $\pm 15$~MeV/$c^2$ of the nominal $D^0$ mass.
For the N4 process, we require that the smaller neutral
$K\pi$ mass  is within
$^{+15}_{-20}$~MeV/$c^2$ of the nominal $D^0$ mass.
For the N5 process, we require that the remaining
$K\pi$ combination, when combined with a $\pi^0$ candidate,
has an invariant mass in the range
$1.83 < M(K^+\pi^-\pi^0) <1.89$~GeV/$c^2$. A $\pi^0$
candidate is any pair of photons in the event that 
fits the $\pi^0\to\gamma\gamma$ hypothesis with $\chi^2 <4$.
If there are multiple $\pi^0$ candidates, we select the one 
that results in $M(K^+\pi^-\pi^0)$ closest to the nominal $D^0$ mass.

For the six-prong processes (N6 and C6), we require exactly
six tracks with particle assignments 
$K^+K^-\pi^+\pi^-\pi^+\pi^-$, where all six pass the looser
and two to four pass the more stringent track criteria that are
indicated by (L) and (S) above, respectively.
For the N6 process, one  combination is required to have
$|\Delta M|_1=|M(K^+\pi^-) - m_{D^0}| <15$~MeV/$c^2$ while
the remaining tracks have
$|\Delta M|_2=|M(K^-\pi^+\pi^-\pi^+) - m_{D^0}| <30$~MeV/$c^2$.
When there are multiple combinations, we take the one with 
the smallest $|\Delta M|_1+|\Delta M|_2$.
For the C6 process, we require 
$|M(K^{\mp}\pi^{\pm}\pi^{\pm}) - m_{D^+}| <30$~MeV/$c^2$
for each of the charge combinations, where
$m_{D^+}$ is the nominal $D^+$ mass.

The invariant mass distributions for $D$-meson
candidates reconstructed
according to the above selection procedure are shown 
in Fig.~1.  For all processes, we require that there
are no extra $\pi^0$ candidates with transverse
momenta larger than 100~MeV/$c$. 
We also apply the following kinematical requirements to reject
initial-state radiation or pseudo-Compton processes (ISR veto). 
The invariant mass constructed from all of the tracks
accepted by the more stringent criteria is less than 4.5~GeV/$c^2$ 
(here a zero rest mass is assigned to each charged track), and 
the  missing mass squared of the system recoiling against the 
detected tracks is larger than  2~(GeV/$c^2)^2$.  
The $D\bar{D}$ candidate system is also required to
satisfy:
$P_z(D\bar{D}) > (M(D\bar{D})^2-49~{\rm GeV^2}/c^4)/(14~{\rm GeV}/c^3) 
+ 0.6~{\rm GeV}/c$,
where $P_z(D\bar{D})$ and $M(D\bar{D})$ are the momentum component in the
$z$ direction in the laboratory frame and the invariant
mass, respectively.
This condition eliminates the ISR events from
$e^+e^- \to D^{(*)}\bar{D}^{(*)} \gamma$ efficiently,
in case the photon is emitted in the forward direction with respect to
the incident electron.
We compute the invariant mass of the $D\bar{D}$ system 
using the measured 3-momenta of each $D$
candidate ($P_D$) and energy determined
from $E_D = \sqrt{P_D^2 + m_D^2}$, where
$m_D$ is the nominal mass of the neutral or charged
$D$.

We calculate $P_t(D\bar{D})$, the
total transverse momentum in the $e^+e^-$
center-of-mass (c.m.) frame with respect to the 
incident $e^+e^-$
axis that approximates the direction of the
two-photon collision axis. 
In the two-dimensional region
$M(D\bar{D})<4.3$~GeV/$c^2$ and
$P_t(D\bar{D}) < 0.2$~GeV/$c$,
we find 159 N4-process events,
110 N5-process events, 240 N6-process events
and 86 C6-process events. 

\begin{figure}
\centering
\includegraphics[width=8.5cm]{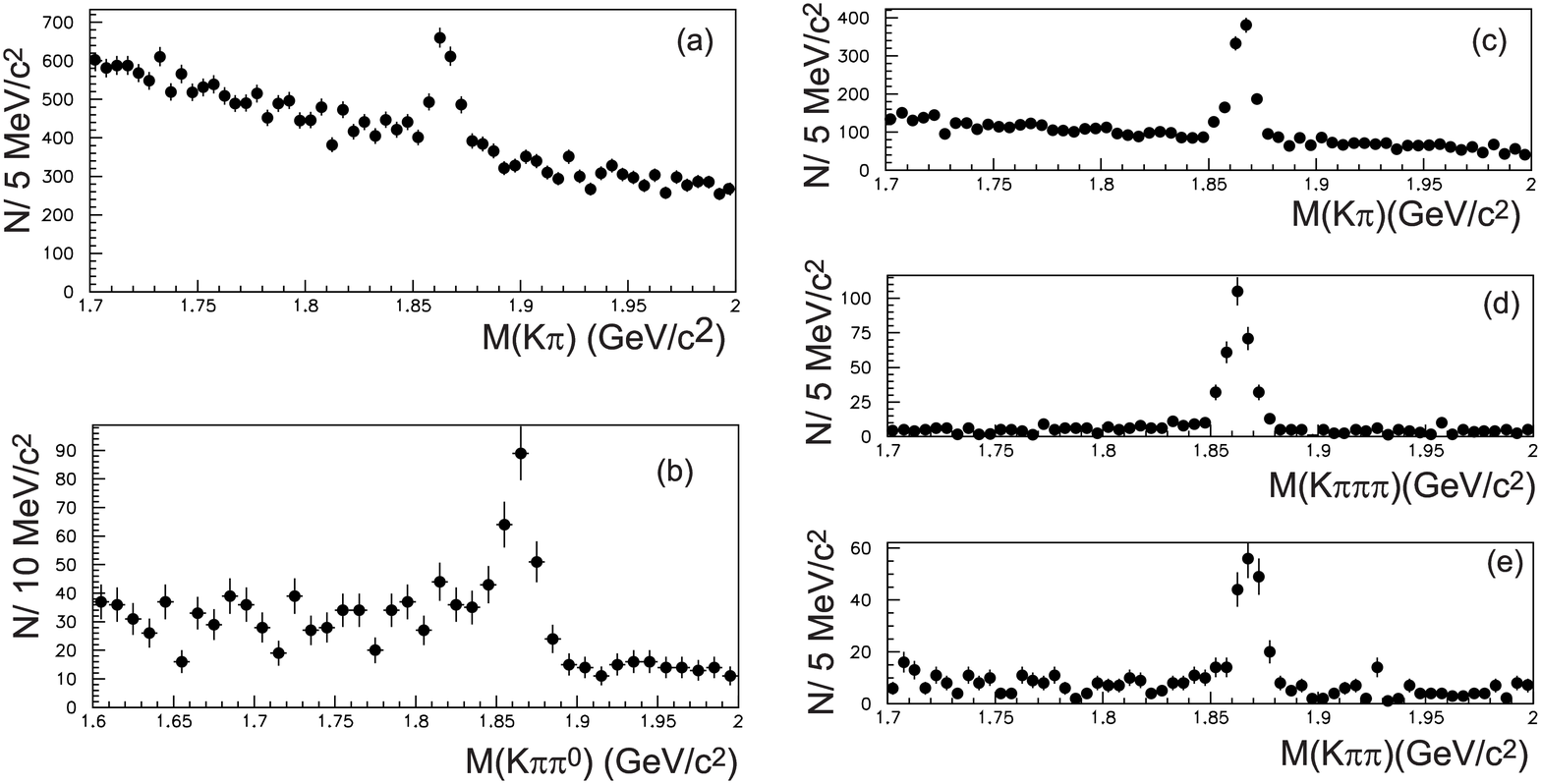}
\label{ddfig1}
\centering
\caption{
{\bf (a)} Invariant mass distribution of $K^{\mp}\pi^{\pm}$
for N4 candidate events. {\bf (b)} Invariant mass 
distribution of $K^+\pi^-\pi^0$ in N5 candidate events
where an accompanying $D^0 \to K^-\pi^+$ (tagged $D$) is found.
{\bf (c)} Invariant mass distribution of $K^{\mp}\pi^{\pm}$
in N6 candidates. 
{\bf (d)} Invariant mass distribution of $K^+\pi^-\pi^+\pi^-$ 
in N6 candidates with a tagged $D^0 \to K^-\pi^+$.
{\bf (e)} Invariant mass distribution of $K^+\pi^-\pi^-$
in C6 candidates with a tagged $D^+ \to K^-\pi^+\pi^+$.
}
\end{figure}


\section{\boldmath Distributions of the $D\bar{D}$ candidates}
In Fig.~2 we show the  $M(D\bar{D})$ distributions
separately  for $D^0\bar{D}^0$ (sum of N4, N5 and N6) (Fig.~2(a)) 
and $D^+D^-$ (Fig.~2(b)) and for the 
combined charged and neutral channels 
(Fig.~2(c)).  In the figures, two event concentrations are evident:
one near 3.80~GeV/$c^2$ and another near 3.93~GeV/$c^2$. 
Here, we have applied the requirement $P_t(D\bar{D}) < 0.05$~GeV/$c$ 
to enhance 
exclusive two-photon $\gamma \gamma \to D\bar{D}$ production.

The invariant-mass distribution for the 
combined $D^0\bar{D}^0$ and $D^+D^-$ channels is shown
for 10-MeV/$c^2$ bins in Fig.~3. The curve is the
result of an unbinned likelihood fit to the data
in the region $3.80< M(D\bar{D})<4.10$~GeV/$c^2$ using
a relativistic Breit-Wigner signal function plus a background 
of the form $\sim M(D\bar{D})^{-\alpha}$,
where $\alpha$ is a free parameter.
The mass dependence of the efficiency and the
two-photon luminosity function is
taken into account in the fit.
These are computed using
the TREPS Monte-Carlo(MC) program~\cite{treps}
for $e^+e^- \to e^+e^-D\bar{D}$ production
together with JETSET7.3 decay routines~\cite{lund73} 
for the $D$ meson decays (using 
PDG2004~\cite{pdg} values for the decay branching
fractions).  The $M(D\bar{D})$  dependence of this product 
is not large.   (At $M(D\bar{D})=3.93$~GeV/$c^2$, 
there is a $\sim 13\%$ decrease in this product
for a  0.1~GeV/$c^2$ increase in $M(D\bar{D})$.)

The results of the fit for the resonance mass, width and
total yield of the resonance are $M=3931 \pm 4 (stat)$~MeV/$c^2$,
$\Gamma = 20 \pm 8 (stat)$~MeV and $41 \pm 11 (stat)$~events, respectively.
The mass resolution, which is estimated by MC to be
2-3~MeV/$c^2$, is neglected in the fit.
The statistical significance of the peak is $5.5\sigma$, which
is derived from the square root of the difference of the
logarithmic-likelihoods for fits with and without a resonance 
peak component, shown in the figure as solid and dashed curves,
respectively.

Systematic errors for the parameters $M$ and $\Gamma$ are
2~MeV/$c^2$ and 3~MeV, respectively. The former
is dominantly due to the uncertainty in the mass of 
the $D$ mesons (1~MeV/$c^2$ for the resonance mass) 
and the choice of the Breit-Wigner function formula (1~MeV/$c^2$).
We consider here several different Breit-Wigner
functional forms for spin 0 and 2 resonances, phase-space
and wave-function variations.
In the latter, we also consider the effects 
of the finite invariant-mass resolution in the fit.

\begin{figure}
\centering
\includegraphics[width=6.5cm]{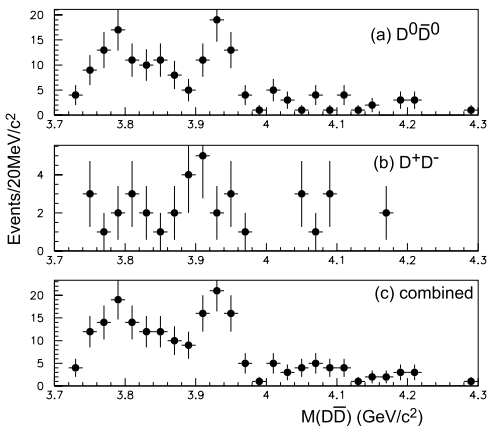}
\label{fig:ddfig4}
\centering
\caption{Invariant mass distributions 
for the {\bf (a)} $D^0\bar{D}^0$ channels 
and {\bf (b)} the $D^+D^-$ mode.  {\bf (c)} The
combined $M(D\bar{D})$ distribution.}
\end{figure}

\begin{figure}
\centering
\includegraphics[width=6.5cm]{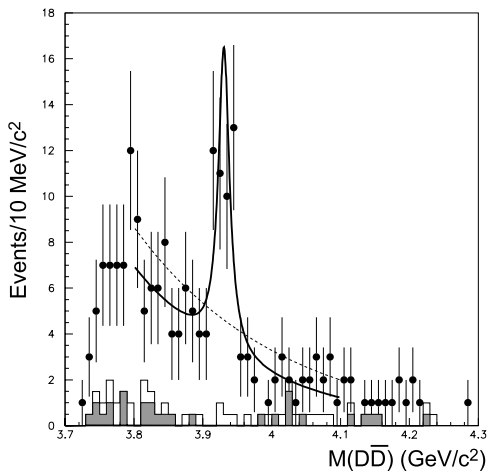}
\label{fig:ddfig5}
\centering
\caption{The sum of the $M(D\bar{D})$
invariant mass distributions for all four processes.
The curves show the fits with (solid) and without 
(dashed) a resonance component.
The histograms show the distribution of the events
from the $D$-mass sidebands (see the text). }
\end{figure}

The $P_t(D\bar{D})$ distribution
in the peak region, $3.91 < M(D\bar{D}) < 3.95$~GeV/$c^2$,
is shown in Fig.~4.  Here the $P_t$ requirement
has been relaxed.  The experimental
data are fitted by a shape that is
expected for exclusive two-photon $D\bar{D}$ production
plus a linear background.  We expect
non-charm and non-exclusive backgrounds to be nearly linear 
in $P_t(D\bar{D})$.  The fit uses a
binned-maximum likelihood method with the scales of the two
components treated as free parameters. The 
linear-background component, 
$0.6 \pm 0.4$ events for $P_t(D\bar{D})
< 0.05$~GeV/$c^2$, and the goodness of 
fit, $\chi^2/d.o.f=18.7/18$, indicate that the events
in the peak region originate primarily from exclusive two-photon
events.

The $P_t(D\bar{D})$ distribution produced
 by $D\bar{D}^*$ and $D^*\bar{D}^*$ events
 is expected to be distorted by the
 transverse momentum of the undetected slow pion(s),
 which peaks around 0.05~GeV/$c$.
 Such a distortion is not evident
 in the observed distribution.
 From a fit that includes a $D^{(*)}\bar{D}^*$
 component, together with $D\bar{D}$ production and a
 linear background, we find that at the 90\% confidence level,
 less than 6 of the 46 events observed in the selected
 $P_t$ region originate from $D^{(*)}\bar{D}^*$.


\begin{figure}
\centering
\includegraphics[width=6cm]{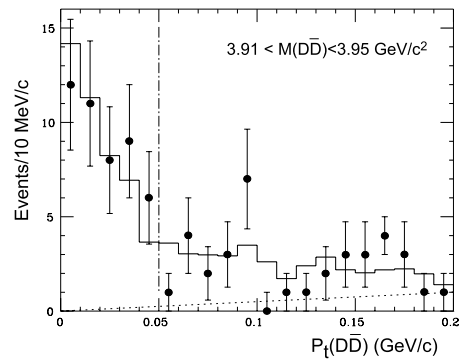}
\label{ddfig6b}
\centering
\caption{The experimental $P_t(D\bar{D})$ distribution 
(points with error bars) for events 
in the $3.91 < M(D\bar{D}) < 3.95$~GeV/$c^2$ region
and the fit (histogram) based on the exclusive 
$\gamma\gamma \to D\bar{D}$
process MC plus a linear background (dotted line). 
The dot-dashed line shows the location of the
$P_t$ selection requirement. }
\end{figure}


  We investigate possible backgrounds
from non-$D\bar{D}$ sources using $D$-sideband events.  
The histograms in Fig.~3
show the invariant mass distributions for events
where the $D$-meson is replaced by a 
hadron system from a 
$D$-signal mass sideband region below
the signal region with the same width as the signal mass region. 
We use two types of sideband events:
one where one $D$-meson candidate is in the signal
mass region (shaded histogram), and another where
both entries are from the sidebands (open-histogram).
Since there is no significant difference between the
two sideband distributions, we conclude that
the sideband events are dominated by non-charm backgrounds.
Since the distributions of the two kinds
of the sideband events are essentially equivalent, we combine
them and scale by a half in order
to compare to the $D\bar{D}$ signal yield. 

The sideband-event distribution has 
no enhancement in the peak region but does include
a broad cluster around 3.80~GeV/$c^2$ 
with a shape that is similar to the
low mass enhancement seen for 
the $D\bar{D}$ candidates.
Since the level of the sideband events is only 
10-20\% of the $D\bar{D}$ candidates, we conclude that 
the lower mass enhancement is dominantly $D\bar{D}$ 
(inclusive or exclusive) events.


Figures~5(a) and (b) show the $M(D\bar{D})$
distributions for events with
$|\cos \theta^*|<0.5 $  and $|\cos \theta^*|>0.5$,
respectively, where
$\theta^*$ is the angle of a $D$ meson relative 
to the beam axis in the $\gamma\gamma$ c.m. frame.
It is apparent that the events in the 3.93~GeV/$c^2$
peak tend to concentrate at small $|\cos \theta^*|$ values.

The points with error bars in Fig.~5(c) show 
the event yields in the 
$3.91~{\rm GeV/}c^2$ to $3.95~{\rm GeV/}c^2$ 
region versus  $|\cos\theta^* |$.
Background, estimated from events
in the $M(D\bar{D})$ sideband, is
indicated by the histogram. 
A MC study indicates that the 
efficiency is uniform in $|\cos\theta^*|$.
For a spin-0 resonance this distribution
should be flat.  In contrast, a spin-2 resonance 
is expected to be produced with helicity-2 along 
the incident axis~\cite{bellechic2,chic2theo}, in
which case the expected  angular distribution
is $\propto \sin^4 \theta^*$.

The solid curve in Fig.~5(c) shows the expectation
using $\sin^4 \theta^*$ to represent the 
signal plus a term proportional to $1 + a\cos^2 \theta^*$ 
that interpolates the  background (dotted  curve).
The comparison to the data has $\chi^2/d.o.f. = 7.3/9.$
Here the functions are normalized
to the total numbers of signal and background events.
(We fix the number of the background events to be 13,
which is obtained from the fit to the invariant-mass
distribution.)
A comparison using a constant term to represent the
signal (dashed curve) gives a much
poorer fit: $\chi^2/d.o.f.= 32.2/9$. 
The data significantly prefer a spin two
assignment over spin zero.

\begin{figure}
\centering
\includegraphics[width=8cm]{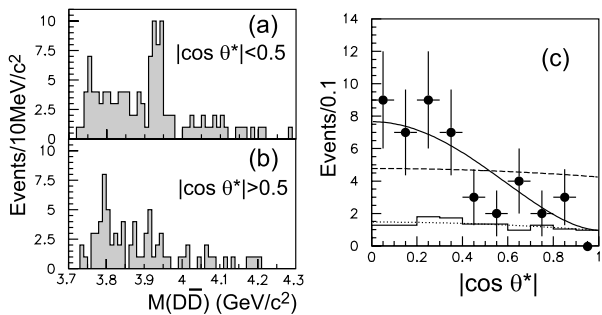}
\label{ddfig7}
\centering
\caption{$M(D\bar{D})$  distributions for
{\bf (a)} $|\cos \theta^*|<0.5 $  and
{\bf (b)} $|\cos \theta^*|>0.5$.
{\bf (c)} The $|\cos \theta^*|$ 
distributions in the $3.91 < M(D\bar{D}) <3.95$~GeV/$c^2$ region
(points with error bars) and background 
scaled from the  $M(D\bar{D})$ sideband (solid histogram). 
The solid and dashed curves are expected distributions  
for the spin=2 (helicity=2) and spin=0 hypotheses,
respectively, plus a curve that interpolates
the non-peak background (dotted), with total area
normalized to the observed number of events. 
}
\end{figure}

\section{Two-photon decay width}
No charmonium state that decays into $D\bar{D}$ with a
mass near 3.93~GeV/$c^2$ has been previously reported.
We find no corresponding event concentration in this 
mass region in the sample of ISR events (normally
rejected by the ISR veto), which would be
expected in case of production of 
a $J^{PC}=1^{--}$ meson ($\psi$).
 

Using the number of observed signal events
and the branching fractions and efficiencies 
for the four decay channels, we determine
the product of the two-photon decay width 
and $D\bar{D}$ branching fraction 
(multiplied by the spin factor) to be
$(2J+1)\Gamma_{\gamma\gamma}(Z(3930)) 
{\cal B}(Z(3930)\to D\bar{D})=1.13 \pm 0.30(stat.)$~keV.
Here, we define 
$ {\cal B}(Z(3930)\to D\bar{D}) ={\cal B}(Z(3930)\to D^0\bar{D}^0)+
{\cal B}(Z(3930)\to D^+D^-)$ and assume 
${\cal B}(Z(3930)\to D^0\bar{D}^0) = 
{\cal B}(Z(3930)\to D^+D^-)$ according to isospin invariance,
where $Z(3930)$ is used as a tentative designation for
the observed state.

The observed signals for the $D^{0}\bar{D}^{0}$ and $D^+D^-$ modes 
are consistent with isospin invariance.
The results on mass, decay angular distributions
and $\Gamma_{\gamma\gamma}$~\cite{muenz} 
are all  consistent with expectations for the 
$\chi'_{c2}$, the $2^3P_2$ charmonium state.

We assign a 16\% total systematic error to the present measurement.
This is primarily due to uncertainties in the track reconstruction 
efficiency (7\%), luminosity function (5\%), MC statistics (7\%) 
and the $D$-meson branching fractions (9\%), added in quadrature.  
Using $J=2$, the above result gives 
$\Gamma_{\gamma\gamma}{\cal B}(Z(3930) \to D\bar{D})=
0.23 \pm 0.06(stat) \pm 0.04(sys)$~keV.

\section{Conclusion}
  We have observed an enhancement in $D\bar{D}$ invariant mass 
near 3.93~GeV/$c^2$ in $\gamma \gamma \to D\bar{D}$  
events.  The statistical significance of the signal
is $5.5\sigma$.
The observed angular distribution 
is consistent with two-photon production of
a tensor meson.
Preliminary results for the mass, width, and the product of
the two-photon decay width times the branching fraction to $D\bar{D}$ 
are: $M=3931 \pm 4(stat) \pm 2(sys)$~MeV/$c^2$, 
$\Gamma = 20 \pm 8(stat) \pm 3(sys)$~MeV
and $\Gamma_{\gamma\gamma}{\cal B}(\to D\bar{D})=0.23 \pm 0.06(stat) \pm 
0.04(sys)$~keV (assuming $J=2$), respectively.  The
measured properties are consistent with expectations
for the previously unseen $\chi'_{c2}$ 
charmonium state.

\ \\
We thank the KEKB group for the excellent operation of the
accelerator, the KEK cryogenics group for the efficient
operation of the solenoid, and the KEK computer group and
the NII for valuable computing and Super-SINET network
support.  We acknowledge support from MEXT and JSPS (Japan);
ARC and DEST (Australia); NSFC (contract No.~10175071,
China); DST (India); the BK21 program of MOEHRD and the CHEP
SRC program of KOSEF (Korea); KBN (contract No.~2P03B 01324,
Poland); MIST (Russia); MHEST (Slovenia);  SNSF (Switzerland); NSC and MOE
(Taiwan); and DOE (USA).


\begin{thebibliography}{99}
\bibitem{godfrey}
S.~Godfrey and N.~Isgur, Phys. Rev. D {\bf 32}, 189 (1985).
\bibitem{eichten}
E.~Eichten, K.~Lane and C.~Quigg, Phys. Rev. D {\bf 69}, 094019 (2004).
\bibitem{belledccbar}
Belle Collaboration, K.~Abe et al., BELLE-CONF-0517, hep-ex/0507019 (2005).
\bibitem{y3940}
Belle Collaboration, S.-K.~Choi et al., Phys. Rev. Lett. {\bf 94}, 182002 (2005).
\bibitem{kekb}
S.~Kurokawa and E.~Kikutani, Nucl. Instr. and. Meth. A {\bf 499}, 1 (2003),
and other papers included in this volume.
\bibitem{belle}
A.~Abashian {\it et al.} (Belle Collab.),
Nucl. Instr. and Meth. A {\bf 479}, 117 (2002).
\bibitem{treps}
S.~Uehara, KEK Report 96-11 (1996).
\bibitem{lund73} H.-U. Bengtsson and T.Sj\"{o}strand, Comp. Phys.
Commun. {\bf 46}, 43 (1987); T.Sj\"{o}strand, CERN preprint, TH-6488-92 (1992).
\bibitem{pdg} Particle Data Group, S.Eidelman et al., Phys. Lett. B {\bf 592},
 1 (2004) (URL: http://pdg.lbl.gov).
\bibitem{bellechic2}
Belle Collaboration, K.~Abe et al., Phys. Lett. B {\bf 540}, 33 (2002).
\bibitem{chic2theo}
M.~Poppe, Int. J. Mod. Phys. A {\bf 1}, 545 (1986);
H.~Krasemann, J.A.M.~Vermaseren, Nucl. Phys. B {\bf 184}, 269 (1981).
\bibitem{muenz}
C.R.~M\"{u}nz, Nucl. Phys. A {\bf 609}, 364 (1996).

\end{thebibliography}
\end{document}